# The use of waveform cross correlation to recover the aftershock sequence of the August 14, 2016 earthquake within Sakhalin Island

Ivan Kitov

Institute for the Geosphere Dynamics, Russian Academy of Sciences

Abstract

The method of waveforms cross correlation (WCC) is used to detect signals from aftershocks of the August 14, 2016 earthquake within Sakhalin Island, which had local magnitude ML = 6.1. Arrivals of regular P- and S-waves detected by the WCC method with various master events at 6 regional stations are associated into a set of seismic events called the cross correlation standard event list (XSEL). We compare the XSEL with the bulletin for the same aftershock sequence compiled in routine seismological processing. The principal advantage of the XSEL is expressed in the increasing number of found seismic events with three or more associated stations, a slight decrease in the magnitude threshold of catalogue completeness, and more accurate location of the epicentres for even the smallest aftershocks. The improved aftershock locations tend to cluster in a narrow zone corresponding to the western board of the Central Sakhalin fault, which defines the boundary between the Okhotsk and the Eurasian (Amur) slabs.

**Introduction**

High seismotectonic activity of Sakhalin Island and its shelf is closely related to the global plate tectonics. Therefore, earthquakes with a large magnitude are possible within the island. The natural Sakhalin seismicity can also interact with the intensive anthropogenic activities related to oil and gas extraction. Identification and quantitative description of the links between these processes is possible with various methods of observation, and continuous seismic monitoring as one of standard techniques. Seismic measurements on a dense network allow revealing temporal, spatial and magnitude distributions of natural and industrial sources.

The tasks of seismic observations conducted by a regional network in a confined area, such as Sakhalin Island and its shelf, include detection of signals from various sources and association of the detected signals into seismic events. Decisive information on seismic events is usually not available from independent sources. Therefore, processing of network seismic data requires the creation of seismic event hypotheses based on information about arrival time, amplitude, slowness, and azimuth measured for regular regional phases - P*, $P_g$, $P_n$, $S_n$, $L_g$. The procedure for creation of an event hypothesis can be very complicated, but it is always based on projection of the arrival times and amplitudes of seismic signals at several stations back to a single physical source. In standard processing, the initial set of hypotheses is built in automatic mode, which are then reviewed and confirmed by analysts.

In routine work of seismic networks, additional information that can also be extracted by methods of quantitative comparison of signals from spatially close events, but such methods are still not fully used. Currently, a new methodology is being developed in seismology – waveform cross correlation, WCC, in which signal similarity plays the most important role. Comparison of real time signals with a template one, for which the source is precisely characterized by location and magnitude, simplifies the process of signal detection and association. With all the diversity of sources and variations in seismic process, a larger part of natural seismicity is concentrated in relatively small areas, such as faults, subducting plates, and swarms. Industrial explosions, like many other types of anthropogenic seismicity, are confined to quarries and mines not exceeding several kilometres in diameter. Long-term observations by modern digital equipment allowed collecting seismic waveforms from spatially close and repetitive sources in many global and regional archives. These historical data make possible effective detection of a template signal in continuous recording using the WCC method, which has been used in seismology since the early 1980s [Geller, Mueller, 1980; Israelsson, 1990]. The overall experience of extensive application of the matched filter (*e.g.*, WCC) technique to large archives of digital seismic records dates back to the early 2000s. Schaff *et al*. [2004] used cross correlation in the double difference method to refine the coordinates of weak earthquakes in California. For the studied area, an increase in the location accuracy by one to two orders of magnitude was reported. This increase is related to the decrease in the influence of the model error in the travel time predictions, which mainly defines the accuracy of absolute location. Further improvement in location accuracy for California was reported in [Shearer *et al*., 2005; Richards *et al*., 2006; Waldhauser and Schaff, 2008].

A similar work on location improvement for a large number of events from a long-term catalogue (1985-2005) was conducted for China in [Schaff and Richards, 2011]. Having processed almost 18,000 sources, the authors found that more than 13% of the signals have correlation coefficient of 0.8, i.e. they are located at distances less than 1 km from each other. The refined epicentres were used to obtain important tectonic information - these events were repeated at a frequency that was not consistent with the standard model of stress accumulation/release.

Waveform cross correlation allows significant reduction in the detection threshold of signals. For example, Gibbons and Ringdal [2006] showed that even one channel of the SPITS seismic group makes it possible to detect signals from recurring sources of natural and artificial origin that are virtually invisible for the beamforming method. The matched filter technique reduces the amplitude detection threshold by an order of magnitude. Schaff obtained a similar result for earthquakes in China [Schaff, 2009]. It should be noted that even in cases where the waveforms become less and less similar with distance between events, the detector based on cross correlation (CC) shows good performance [Schaff, 2010]. Schaff and Waldhauser [2010] confirmed this result by examining the catalogue of earthquakes in the Parkfield seismic zone. For global and regional seismic networks, the CC-detector is useful for events at distances up to several hundreds of kilometres [Bobrov *et al*., 2013ab; Adushkin *et al*., 2015].

An important advantage of the cross correlation method is the possibility of obtaining more accurate estimates of the relative magnitude for a set of seismic events. Gibbons and Ringdal [2006] introduced the concept of the size of a slave event with respect to a master event (ME), as the ratio of the RMS amplitudes in the cross correlation window multiplied by the cross correlation coefficient. The relative magnitude is introduced as a logarithm of the relative size [Schaff and Richards, 2011] and has an important property of a lower standard deviation of the magnitude estimate than the estimate obtained by the standard method. Bobrov and co-workers [Bobrov *et al*., 2014] pointed out the drawbacks of such a formula.

When the distance between events increases, the correlation coefficient decreases without any fall in the slave amplitude. The relative magnitude should be based only on the ratio of the RMS amplitudes and only for signals detected by the cross correlation detector. Schaff and Richards [2014] carried out a detailed analysis of both definitions of relative magnitude at a seismic network in China and concluded that the latter definition provides a more accurate and unbiased estimate of the relative size.

One of promising areas for the use of waveform cross correlation is automatic recovery of aftershock sequences, i.e. the events close in time and space as well as in source mechanism. For a regional/global network, tens and hundreds of aftershocks per day are a challenge for interactive processing. Such sequences can occur even after earthquakes with magnitude 5 in case the detection threshold is of 2 to 3 magnitude units. For example, the WCC method detected at regional stations an aftershock with magnitude of 2.2 after the underground explosion conducted in the DPRK on September 9, 2016 [Adushkin *et al.*, 2017]. Harris and Dodge [2011] proposed and tested an automatic system for aftershock detection and grouping. Their work is based on the results obtained with the matched filter in [Nadeau and McEvilly, 1997; Schaff and Richards, 2004; Schaff and Waldhauser, 2005; Schaff, 2009]. In automatic mode, one can recover aftershock sequences from small [Bobrov *et al.*, 2014] to very large [Bobrov *et al.*, 2013a; Bobrov *et al.*, 2015ab] earthquakes.

In this paper, we recover the aftershock sequence of the August 14, 2016 earthquake on Sakhalin Island. It is assumed that not all aftershocks, which can be found using the WCC method, are in the catalogue for this earthquake published by the specialized automated information service EQAlert.ru (http://www.eqalert.ru). The main task is to extend the published catalogue with reliable aftershocks.

**Recovery of the aftershock sequence**

On August 14, 2016 at 11:15:13.1 (UTC) an earthquake occurred in Sakhalin Island with ML = 6.1 according to local data. The earthquake coordinates were estimated using near-regional stations: 50.351° N, 142.395° E. This location is close to that estimated by the International Data Centre (IDC), which is also available at the International Seismological Centre: 50.424° N, 142.381° E. The IDC origin time 11:15:12.62 (UTC) is slightly different likely due to the depth fixed to zero, while the local measurements put the hypocentre to 9 km depth. The estimates of body wave magnitude vary from 5.4 (IDC) to 5.9 (GFZ Potsdam). Several hundreds of aftershocks can accompany such an intermediate magnitude earthquake within two to three weeks.

Figure 1 shows positions of 6 three-component seismic stations (A732, A759, ARGI, LNSK, NGLK and NYSH) relative to the aftershock zone. Distances and azimuths to the epicentre of the main shock are given in Table 1. The nearest station ARGI is located at 112 km, which may result in the highest detection rate. Station A759 is located within Sakhalin, about 2 km from the coastline. Station A732 is on the continent and the sea leg can result in additional attenuation and thus reduce the number of detectable signals, both by conventional and the WCC-based methods. Despite the larger distance from the aftershock zone, stations A759 and A732 play a key role in the aftershock location, as they are on the opposite side compared to the other 4 stations. The main event was recorded at several other stations of the Sakhalin network, but they make a minimal contribution to the detection of the weakest aftershocks. Therefore, processing of data from these stations is impractical.

Table 1. Distances and azimuths to stations

| Station | A732 | A759 | ARGI | LNSK | NGLK | NYSH |
|---|---|---|---|---|---|---|
| *Distance, km* | 208 | 144 | 112 | 134 | 167 | 135 |
| *Azimuth, deg.* | 229 | 190 | 12 | 29 | 18 | 11 |

The aftershock catalogue from August 14 to August 31, 2016 was published by the EQAlert.ru service and includes 134 events. One of them is far from the zone of aftershocks (see Figure 1). We use this event to evaluate the reliability of the WCC detection method - no master events from the actual aftershocks should find it. In Figure 2, we show the distribution of aftershocks relative to the main shock. The maximum linear size of the aftershock zone is approximately 40 km and it is aligned along the southeast to the northwest direction.

Figure 3a depicts the frequency distribution of the aftershock magnitudes, which has a peak between 2.5 and 3.0. The catalogue completeness is limited to magnitude 2.75-3. Since the WCC method is used to find more events, the magnitude threshold of catalogue completeness has to be lowered. The depth distribution of the Sakhalin aftershocks peaks between 10 and 15 km as shown in Figure 3b: all aftershocks are within the crust. Since the method of local association, which we describe later in this paper, uses the source to station travel time, we have to take into account the depth of master events when calculating the origin time for a source fixed to the free surface. The correction is calculated as the travel time from the depth of the master event to the free surface for the known velocity structure of the crust. For the deepest aftershock (44 km), the travel time correction is almost 7 seconds.

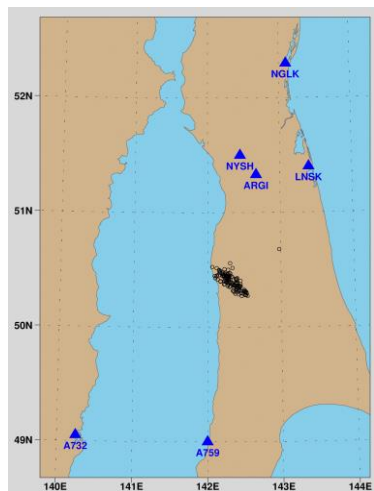

Fig. 1. Six closest seismic stations and the aftershock area

**Selection of waveform templates**

All measured signals associated with seismic sources from the catalogue can serve as waveform templates. The principal requirements for the signals are as follows: high signal-to-noise ratio and representativeness, understood as similarity to a majority of signals generated by events in the same zone. The latter condition cannot be satisfied before calculating cross correlation coefficients (CC), since these coefficients serve as a measure of similarity. The first condition is easy to fulfil. At first, the signals from larger aftershocks with high SNR are tested. We selected the events from the EQAlert.ru catalogue with magnitudes above 3 detected by at least 5 stations. For all pairwise combinations of these master events, we calculated CC-values at all available stations. In total, 1020 templates were tested of which only 80 from 7 events were selected for further analysis. In Figure 2, these events are highlighted in red and named by their serial numbers in the EQAlert.ru catalogue. The selected master events (MEs) almost uniformly cover the entire zone of aftershocks with an emphasis on areas of the highest concentration. The distance to the nearest master event does not exceed 10 km. In Figure 4, waveforms from the main shock are displayed. At all 6 stations, clear P- and S-waves are observed.

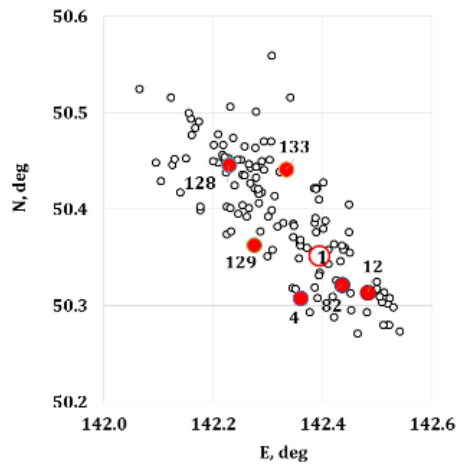

Fig. 2. Spatial distribution of the aftershocks from the published catalogue. As master events, we use the main shock (1) and 6 larger aftershocks covering the whole area of post-seismic activity.

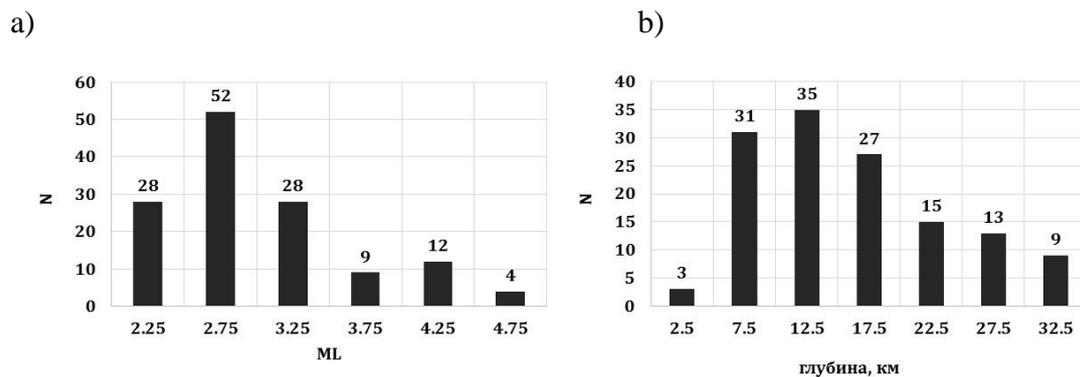

Fig. 3. a) Frequency distribution of the aftershock magnitudes (ML) with a peak between 2.5 and 3. The main event is not shown. b) Frequency distribution of depths with a peak between 10 and 15 km.

For each ME, two templates are created at a given station - for the P- and S-waves separately. To distinguish between these templates, we add a phase-name to the station name: ARGIP or A732S. Regular regional phases P and S are detected independently and can create not correlating templates. For earthquakes, the S-wave amplitude can significantly exceed the amplitude of wave P, both at the nearest and remote stations. Therefore, we allow an S-wave arrival to be associated with a source in the absence of P-wave arrival. This leads to the possibility of creating hypotheses of events consisting only of S-waves detected by the WCC method. Formally, the P and S signals create two virtual stations for phase association with the same coordinates.

Before we start signal detection by standard method, all waveforms are filtered in 7 frequency bands by the 3-rd order Butterworth filter: F1: 1- 3 Hz, F2: 2-4 Hz, F3: 3-6 Hz, F4: 4-8 Hz, F5: 6- 12 Hz, F6: 8-16 Hz and F7: 12-24 Hz. When the ratio of a short-term-average and long-term-average (STA/LTA) is above some predefined threshold on at least one filter, a signal is detected. The largest SNR on the vertical channel at a given time is selected among 7 filters. The duration of the STA window is 0.5 s, and for the LTA - 60 s. All templates are composed of three-component records filtered in the same ranges. In Figure 4, all three components are shown: Z, N and E for 4 filters - from F2 to F5. Each template contains 10 s of ambient noise before the signal, and the duration of the template is 60 s. The P-wave template includes the S-wave, which arrives 15 to 20 s later. The noise before the signal is excluded from the CC calculations in all cases.

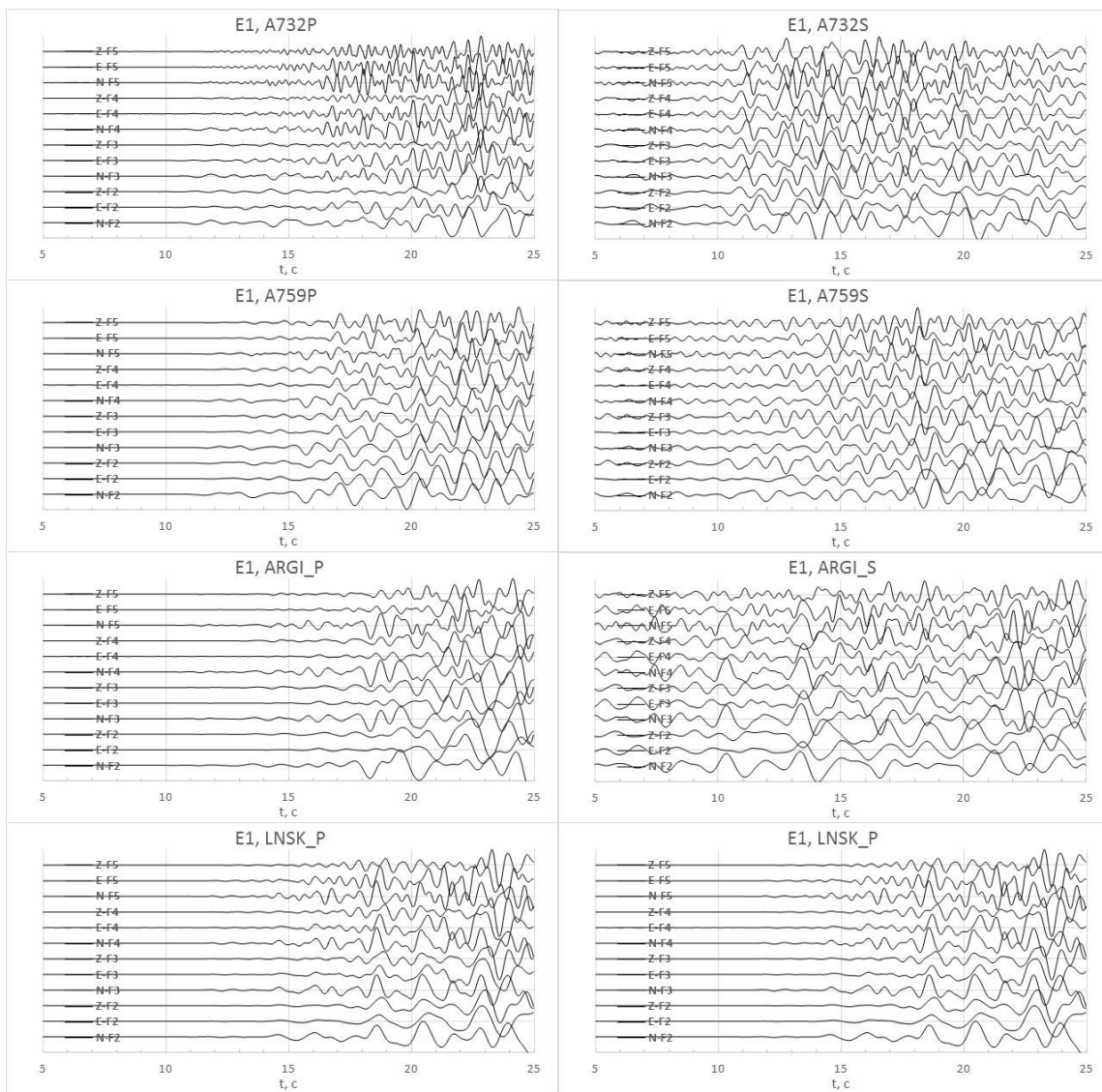

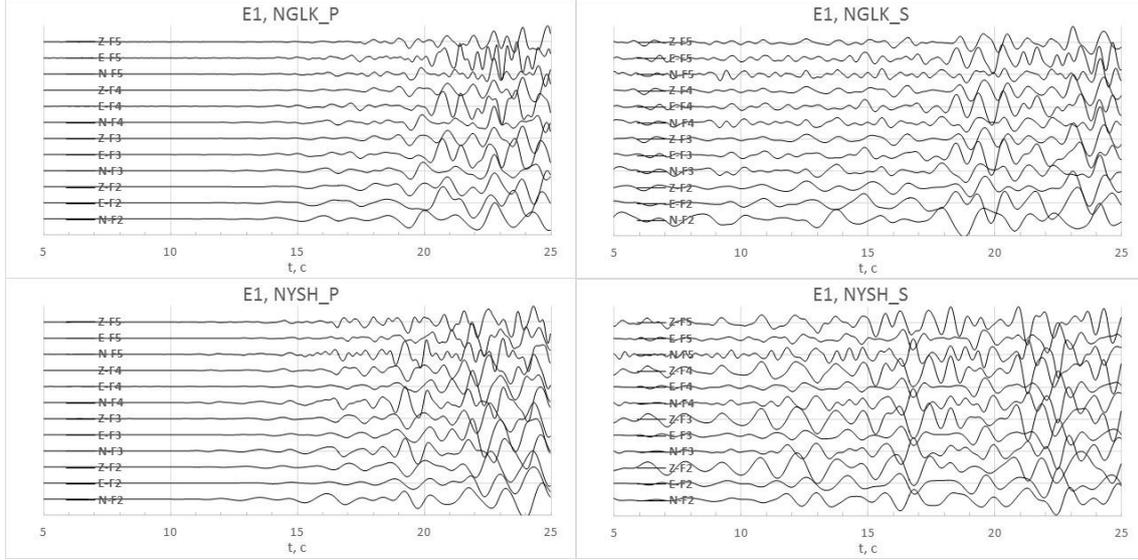

Figure 4. Waveforms of the P-wave and S-wave from the main shock (E1) at 6 stations. Templates include 10 seconds before the signals. For the S-waves, the segment in front of the signal is the P-wave coda.

The cross correlation coefficient, CC, is calculated for each channel, $j$, of a three-component seismic station. The waveform templates have a length of N consecutive samples for each channel, $m_{jn}(t_0)$. The total number of samples in a multi-channel template is determined as a product of the length of the template and the number of channels. For a discrete recording, on each channel, $u_j(t)$, we determine the cross correlation coefficient, $CC_j$, for the absolute time $t$:

$$CC_j(t) = m_{jn}(t_0) \cdot u_{jn}(t) / (\|m_{jn}(t_0)\| \cdot \|u_{jn}(t)\|) \quad (1)$$

where $\| \cdot \|$ determines the L2-norm for the corresponding time series of length N. The aggregated (station) correlation coefficient is calculated by averaging over M = 3 channels:

$$CC(t) = \sum_{j=1}^{M} CC_j(t)/M \quad (2)$$

The resulting trace $CC(t)$ has the same sampling rate as the original record, and the same duration, except for the last segment of length N. The three-component trace is converted into a one-component $CC$ time series. We also have an option of calculating $CC$ only for the vertical component. To detect signals, including those below the noise level, we use the same detector based on the STA/LTA ratio. For the cross correlation trace, the signal-to-noise ratio is denoted $SNR_{CC}$. In standard mode, we use the same values LTA=60 s, STA=0.5 s as for the original waveforms. An important parameter that determines the success of the WCC method is the template duration. For the P-wave, the arrival of the wave S can be considered a natural restriction. Therefore, the duration of the template P for all stations is 8 s. For the S-wave, the arrival of surface waves in several tens of seconds becomes a limitation. We fix the minimum possible value of 10 s for all stations.

Out of seven $SNR_{CC}$ values (for 7 filters) for each time sample, we choose the largest. If this maximum value exceeds the detection threshold, an automatic search of the maximum $SNR_{CC}$ value for all filters is performed for the whole length of the template after the first CC-arrival.

The time of actual arrival should be near the $SNR_{CC}$ maximum, *i.e.* near the point where the slave signal is best matched by the template. Since STA and LTA are running averages, they smooth out the peaks in the *CC* time series and can introduce some bias into the estimates of exact arrival time. Therefore, we find the maximum of the absolute value | CC | within ± 2 s from the maximum $SNR_{CC}$. The arrival time corresponds to the peak | CC |. In automatic processing, sequential signals cannot be closer than 20 seconds from each other.

**Results of signal detection using cross correlation**

Using the template signals, we calculate continuous CC-traces for three individual channels, averaging them to obtain the aggregated CC-trace for each station, and then apply the signal detection procedure using the $SNR_{CC}$ threshold. Figure 5 shows several examples of detection. Figure 5a presents the *CC* and $SNR_{CC}$ curves for autocorrelation of the main event at station A732 (A732-P). The *CC* starts to grow when the tail of the 8-second template reaches the sought signal on the continuous seismogram. The CC trace oscillates 8 s before it reaches the peak CC=1 when the template coincides with itself. The $SNR_{CC}$ overcomes the threshold of 3.5 when the *CC* starts to grow, but this is not considered as a signal arrival since the maximum $SNR_{CC}$ can be somewhere within the length of the correlation window. After finding the maximum $SNR_{CC}$, the search for the exact arrival time continues since the peak *CC* gives a more reliable value. Near the $SNR_{CC}$ threshold, the influence of microseismic noise on the estimates of STA and LTA increases, which may bias the arrival time estimate. In turn, errors in the arrival time affect the effectiveness of phase association into event hypotheses. Figure 5b shows a similar autocorrelation example for the S-wave at LNSK (LNSK-S).

Figure 5c shows an example of clear P-wave arrival at station A732. The sharp $SNR_{CC}$ peak is much higher than the threshold value (3.5) and allows to accurately estimating the onset time. In Figure 5d, we present a detection example for a weak S-wave at LNSK. The $SNR_{CC}$ is just above the threshold, but the corresponding peak is sharp enough and well above the background values related to the microseismic noise. Based on the properties of $SNR_{CC}$ in cases c) and d), one may suggest that the detection threshold is somewhat overestimated. When the threshold is lower, the number of detectable signals from real aftershocks might increase. However, the false alarm rate also increases. A finer tuning of the threshold value for the CC-detector is a separate task, however. Solution of such a problem is likely possible only with the accumulation of sufficient information about the properties of microseismic noise.

For two signals from spatially close events it is instructive to introduce a parameter that determines both the relative amplitude of two signals and the relative magnitude of two events. The ratio of the signal norms: | **x** | / | **y** |, where **x** and **y** are the vector signals of the master and slave, respectively [Bobrov *et al.*, 2014]. The logarithm of the ratio, determines the magnitude difference between two events or the relative magnitude:

$$dRM = \log(|\mathbf{x}|/|\mathbf{y}|) = \log|\mathbf{x}| - \log|\mathbf{y}| \qquad (3)$$

This difference has a clear physical meaning for close events with similar signals. In addition to estimating the magnitude of the slave event, the relative magnitude is a reliable dynamic parameter improving the reliability of phase association at several stations. The meaning of the dynamic matching of the arrivals is that the deviation of the station relative magnitude from the network should not be out of some narrow tolerance range. In this study, the tolerance range is ±0.5 units of magnitude.

The dRM calculations are performed in the frequency band where the signal has been detected, and the length of the vectors **x** and **y** is exactly equal to the length of the template. When there are many master events, the relative magnitude estimates can be carried out using the LSQ methods applied to all pairs of events. At the same time, for an aftershock sequence, there is always the main shock with an accurately estimated magnitude. Therefore, instead of all ME-slave pairs, one can use the magnitude relative to the main earthquake. The results of dRM calculation at 6 stations throughout the aftershock sequence, including newly detected events, are presented in Appendix 1.

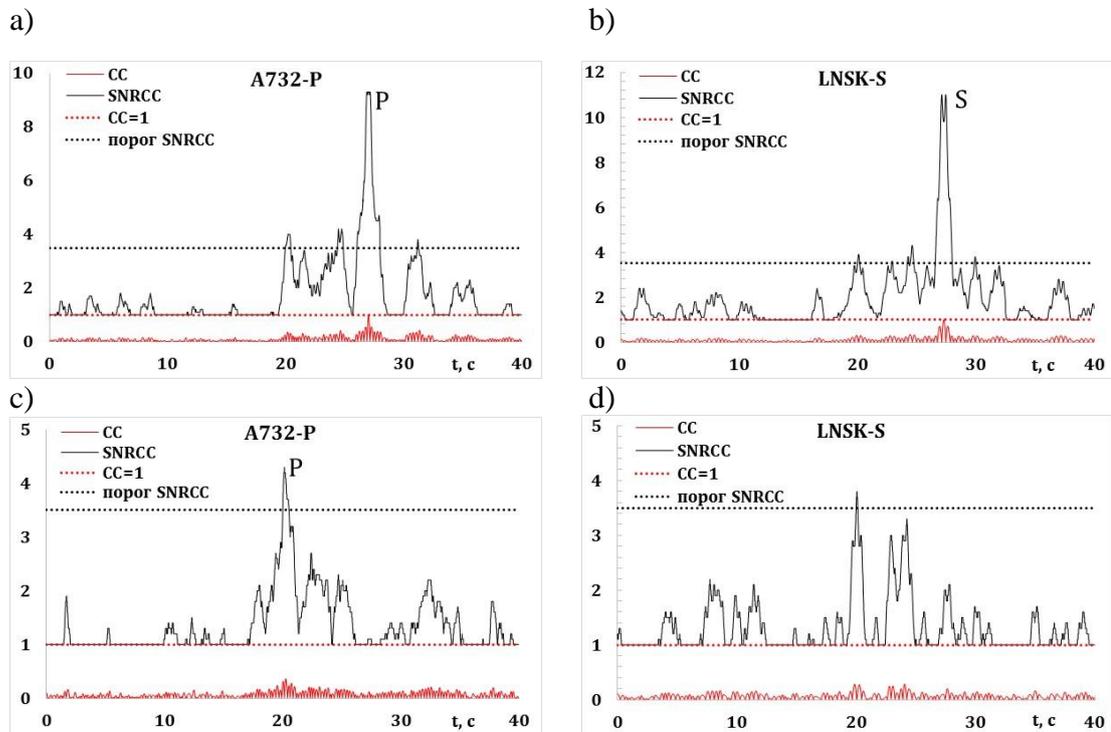

Figure 5. Examples of signal detection using waveform cross correlation. Selected segments of the $CC$ and $SNR_{CC}$ traces are shown. Two cases of autocorrelation, a) and b), illustrate the difference in the procedure for detecting conventional and cross correlation signals - the first exceeding of the threshold $SNR_{CC}$ does not mean a physical signal arrival. Panels c) and d) demonstrate detection of weak P- and S-waves.

**Phase association into seismic event hypotheses**

At 6 stations of the Sakhalin network, we have templates for each ME. Our task is to detect, create hypotheses for and to estimate principal parameters of all events, which match a few quality criteria. In this paper, we use standard, but rather flexible, definition based on the number of stations and the quality of the signals, *e.g.*, $SNR_{CC}$. In routine seismological practice, an event exists if there are signals at three or more stations. Here, different phases and the possibility of azimuth and slowness measurements are not taken into account. However, the use of the WCC method allows more effective use of information on the shape of detected signals. The P- and S-wave from the same source practically do not correlate. This effect increases the efficiency of the WCC method: it allows not only not reduction in detection threshold, but also improves signal discrimination. Therefore, the requirement for the number of stations can be eased: the P- and S templates effectively create two stations with the same name. We found that the minimum number of virtual stations for creating a reliable seismic event hypothesis is 4. This means that the minimum number of actual seismic stations is 2 when P- and S-waves are detected at these two stations. We also analyse the change in the number and quality of created events with 5 and 6 stations. Any further increase in the number of associated stations does not give a significant advantage in the reliability of created hypotheses, but significantly reduces the number of events found.

After the detection process is finished, we obtain a set of signals with their arrival times at each station, $t_{ij}$, where $i$ is the detection number at station $j$. For events close to MEs, the travel time to the corresponding station, $tt_j$, can be accurately represented by the theoretical ME/station travel time as well as the empirical travel time, i.e. by the difference of the arrival and origin times. In turn, using the empirical travel time, $tt_j$, obtained from the EQAlert.ru catalogue, and the measured arrival times it is possible to calculate the origin time, $ot_{ji}$ for all detections:

$$ot_{ji} = t_{ij} - tt_j \qquad (4)$$

The set of arrival times at 6 stations is converted into a set of origin times for an unknown number of seismic events. In accordance with one of quality criteria, we assume that any event found by cross correlation should have four or more stations for which the calculated origin times differ by no more than a few seconds. We allow scattering up to 4 s, but the events in the cross correlation catalogue usually have origin time residuals of tenths and hundredths of a second. This condition is equivalent to matching the arrival time residuals in standard location procedure. To determine the absolute origin time for a given event hypothesis, we average the station estimates.

The slave events is supposed to be near one of the MEs. When a slave event is 10 km or more away from the closest ME the travel time from this ME is no longer a good approximation in (4). The procedure for the arrival time reduction to the source should take into account such changes and we introduced an additional set of possible epicentres for the slave events around the MEs. These additional epicentres are evenly distributed over a rectangular grid as shown in Figure 6. The grid size and spacing between nodes are determined by the distance between the MEs. As a given grid increases in size, it may overlap with the grid from one or more neighbouring master events. In this case, there may be a conflict between two master events competing for the same arrival.

For each node, the origin time is calculated as a correction to the empirical travel time. For $k$-th node the correction, $dt_k$, is calculated as the scalar product of the horizontal slowness vector, **S**, for corresponding seismic phase and the ME-node vector, $\mathbf{d_k}$:

$$dt_k = \mathbf{S} \cdot \mathbf{d_k} \tag{5}$$

The slowness is determined by velocities of the P- and S-waves. We selected standard values from the IASP91 velocity model. The more accurate values can be obtained from the regional travel time curves. Now, one can rewrite (4) in the form

$$ot^k_{ji} = t_{ij} - tt_j + dt_k \tag{6}$$

The search over all nodes ensures the selection of the minimum LSQ time residual error in the source. It is very likely that this node is closest to actual slave location. It should be noted that the same arrival times are used for all nodes and the ME-slave cross correlation coefficients remain the same. Each node plays a role of virtual master event - the original ME location effectively moves to another point in space.

Formally, one can move any ME by longer distances. Such a ME can be a "Grand Master" event (GM) [Bobrov *et al*., 2015a], *i.e.* the event replacing those MEs which are closer to the slave. We might select the best (e.g. more stations, higher SNRs, larger representativeness) of all master events for the studied zone, and expand the regular grid for this event to the entire aftershock zone. This would be the best GM for the aftershock zone.

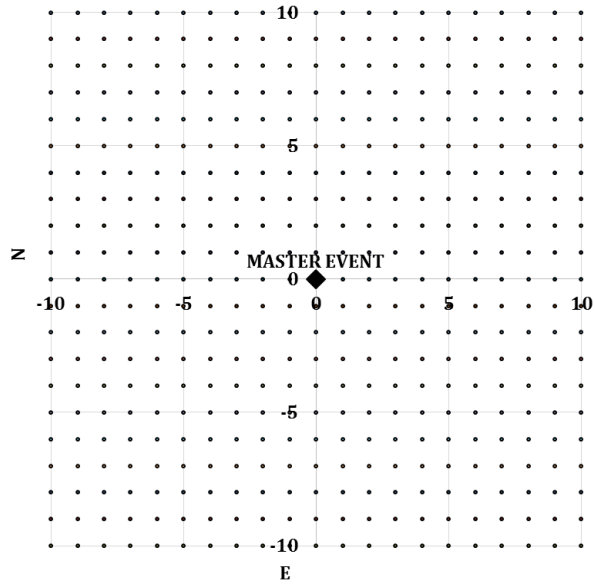

Fig. 6. An example of a grid for slave event location relative to the master event in the centre.

We refer to the origin time alignment process as the local association (LA) in accordance with the name of the global association (GA), often used in standard seismic data processing [Coyne *et al*., 2012]. Indeed, the phases at different stations should be associated only with seismic events close to the MEs. This association procedure should not be sensitive to events outside the radius of cross correlation between MEs and slaves since there should be no CC-detections beyond this radius. Moreover, when a remote event (say, 20 to 30 km) shows high cross correlation coefficients at several stations, the arrival (and thus the origin) times at these stations are scattered far beyond the pre-determined limits (4 s) of the allowed residuals. No event hypotheses can be created.

Operationally, we began the LA process allowing the P-wave origin time residuals of 10 s. The 10 s width approximately corresponds to the travel time difference between a ME in the centre and a slave event located on the border of the aftershock zone. For the ME-slave distance of 10-15 km and regional P-wave slowness of 0.18 s / km, the difference in travel times is 2 to 3 seconds. For two stations in opposite directions from the ME, the difference in the origin time estimates is 4 to 6 s. This is the worst case scenario. The precision of arrival time estimates with the WCC method is of 1 s, which can add 2 s to the origin time difference. For the S-waves, the origin time residuals are approximately two times larger.

For the Sakhalin aftershock sequence, one can assume that the most conservative length of the time segment, in which the origin times obtained from the P- and S-wave arrivals may create an event hypothesis, is ~20 s. When arrival times at several stations belong to the same physical event, their uncorrected origin times have to be within this 20 s segment. Since we have calculated and aligned all origin times for all MEs, the LA procedure sequentially moves to the next origin time and checks for presence of at least 3 more origin times from different stations within the next 20 seconds. When an initial event hypothesis is created with 4 or more stations, we start the grid search to minimize the RMS origin time residuals, and thus, locate the slave event. Individual origin time residuals should not exceed the tolerance limit. When 4 or more origin times do not cluster within the predefined range the initial event hypothesis fails. For a successful event hypothesis, the origin time is calculated as the mean origin time.

Figure 7 depicts 4 examples of event location. The location procedure is based on the distribution of the RMS origin time error and the number of stations. The colour bar to the right of each panel shows the RMS origin time error in seconds. A dense grid of 100x100 nodes was used with a step of 100 m, i.e. 5,000 m in each direction from the centre. The main shock is used as a ME and placed in the centre of the grid. The distribution is an ellipse with the semi-major axis almost perpendicular to the line connecting the two groups of stations to the northeast and southwest of the aftershock zone. It is obvious that in this direction the location accuracy is poor – moving the source in this direction has almost no impact on the origin times.

Figure 7a shows an example of very close ME and slave. The distance between them is approximately 100 m, while the minimum RMS origin time error is a few hundredths of a second. In Figure 7b we present location of an event at a distance of more than 2 km from the ME. In the upper right corner, one can observe a small transition zone to a smaller number of associated stations. With fewer associated stations, the RMS residual becomes smaller, which is reflected in the appearance of a small bright green segment. Figure 7c illustrates a slave location north to the ME and Figure 7d shows an example of a less reliable location, which is expressed in a large RMS residual of ~0.6 s. The ellipse defined by a 0.6 s residual line is also much larger than in the other three panels. This effect is associated with fewer associated stations. To optimize the LA process and relater relative location one can vary the grid size and spacing, the minimum number of associated stations, the length of the association window, and the maximum deviation from the average origin time and obtain different sets of event hypotheses. We have compared the cross correlation catalogues with different minimum number of associated stations with the EQAlert.ru catalogue.

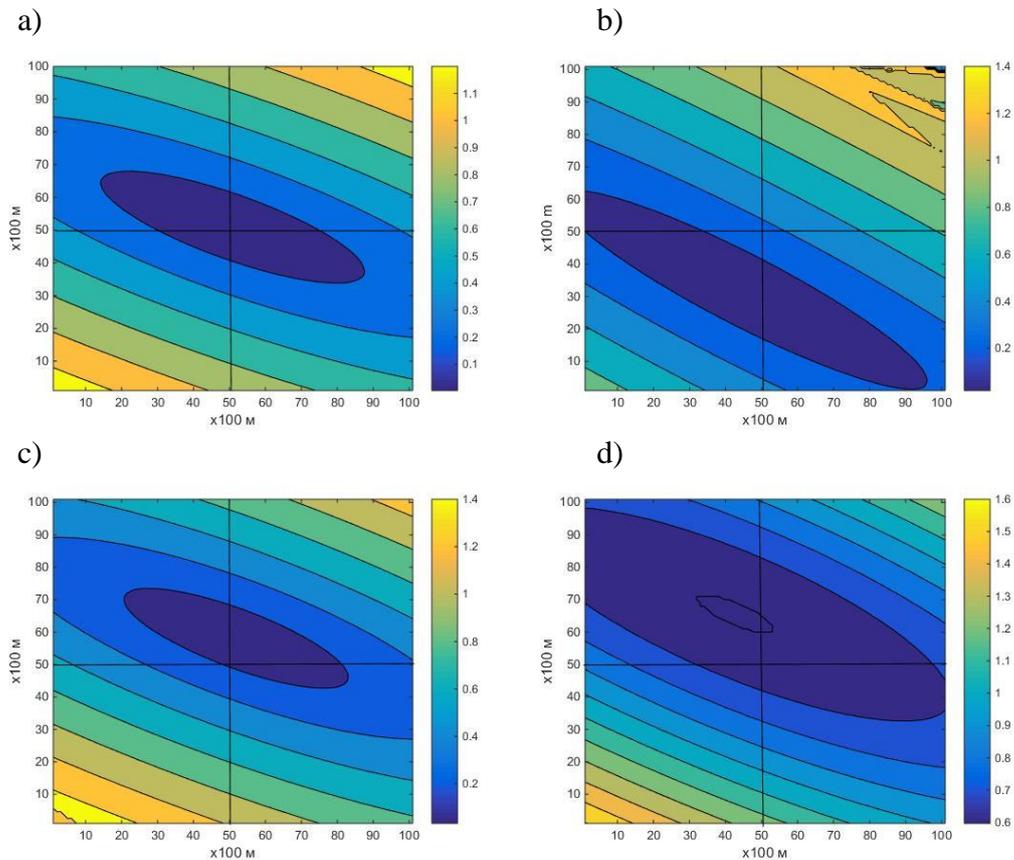

Figure 7. Examples of the RMS origin time residual distribution. The colour bar to the right shows the time residual in seconds. A grid of 100x100 nodes with a centre at the master event location has a step of 100 m. a) Close slave and ME locations. b) and c) ME-slave distance of one to two kilometres with a low scattering in origin times. d) Distance of two kilometres with a large origin time scattering - 0.6 s in the centre of the ellipse.

**Conflict resolution**

Each ME has neighbouring MEs and their templates can well correlate with signals from the same slave event. As a result, very similar event hypotheses are created and may conflict with each other for the same physical seismic signals at several stations. Therefore, a procedure is needed that selects one most reliable hypothesis of the event for each arrival.

The primary selection of the best hypothesis claiming the same arrival is based on the number of associated stations. When only one event hypothesis has the largest number of stations, it is saved as an event of an automatic cross correlation standard event list, XSEL [Bobrov *et al*., 2013ab; Bobrov *et al*., 2015ab]. When several hypotheses have the largest number of stations, we choose the one that has the smallest RMS origin time residual. By definition, this event hypothesis is the most probable, and thus, saved in the XSEL. This stage of the IA process is defined as conflict resolution (CR). From all sets of hypotheses associated with the MEs, the CR creates a single set of events.

After resolving conflicts related to all arrivals, there are two possibilities for the hypotheses losing one or more arrivals. When a given hypothesis violates the minimum number criterion after it loses one or more associated arrivals it has to be rejected. When a sufficient number of associated stations is retained in the hypothesis, it can be saved for further analysis. This approach is important for large areas with many MEs, for example, the aftershock zones of very large earthquakes. Signals from simultaneous events at different sides of such zones can come to some stations with a slight delay, while at other stations the corresponding signals are well separated. In this case, to reject the event hypothesis losing one or a few associated phases would be not correct. For intensive aftershock sequences and many stations, the number of phases lost by an event hypothesis can be large, but the final number of phases may still satisfy the event definition.

For small aftershock zones with just tens of events per day measured at several stations, the probability of an event to be valid when it loses one or two associated phases is low. Therefore, we introduce an additional parameter that determines the number of lost phases for an event hypothesis to be discarded. The number of aftershocks in the EQAlert.ru catalogue in two weeks after the main shock is 133, the number of (virtual) stations is 12, the minimum number of associated stations is 4, and the linear size of the aftershock zone is of 50 km. Considering these figures, we reject all hypotheses losing one or more phases. This decision was based on the fact that two events never occurred simultaneously within the aftershock zone.

**Results of the local association**

To illustrate the results of local association and relative location, we first present the bulletin obtained with the main earthquake used as a ME. Figure 8 shows the origins times for August 14, 2016, which were obtained by back projection of the corresponding arrival times. The y-axis represents the $SNR_{CC}$ estimates for each arrival. In Figure 8, the origin times are obtained for the master event position. Nevertheless, the origin times are well synchronized. The main event has all 12 virtual stations associated.

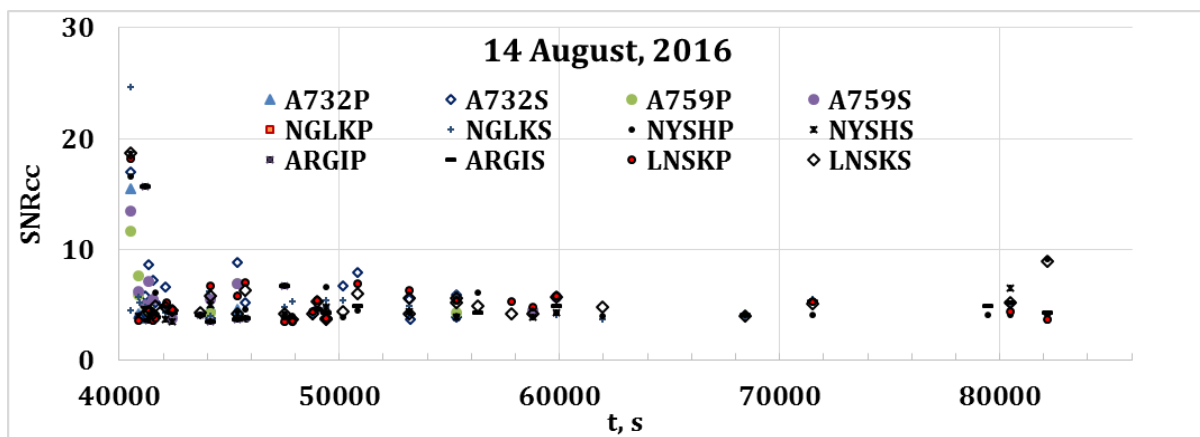

Figure 8. Schematic representation of association of cross correlation detections into seismic events. Several origin times within a few seconds create an event hypothesis.

Right after the main event, the aftershock activity was high. Finer details of this activity are better seen in Figure 9, where the period between 40,000 s and 50,000 s of August 14, 2016 is selected. We compare the origin times for the events from the interactive catalogue EQAlert.ru (vertical dotted line) and the XSEL for the main shock as ME. All origin times of the XSEL events in Figure 9 belong to the final locations. For each event from the interactive catalogue, cross correlation found from 4 to 12 P- and/or S-wave arrivals. The stations and phases forming the XSEL events are shown by different symbols. For the main shock, the detections are obtained by autocorrelation. In general, Figure 9 illustrates the result of the WCC method applied to continuous seismograms at 6 stations. It is important that one ME is able to detect all events from the interactive catalogue within the aftershock zone in the first three hours. This is an example of using the "Grand Master" approach to recovering the aftershock sequence. With six additional aftershocks as MEs (see Figure 2) one can expect a significant increase the number of aftershocks found.

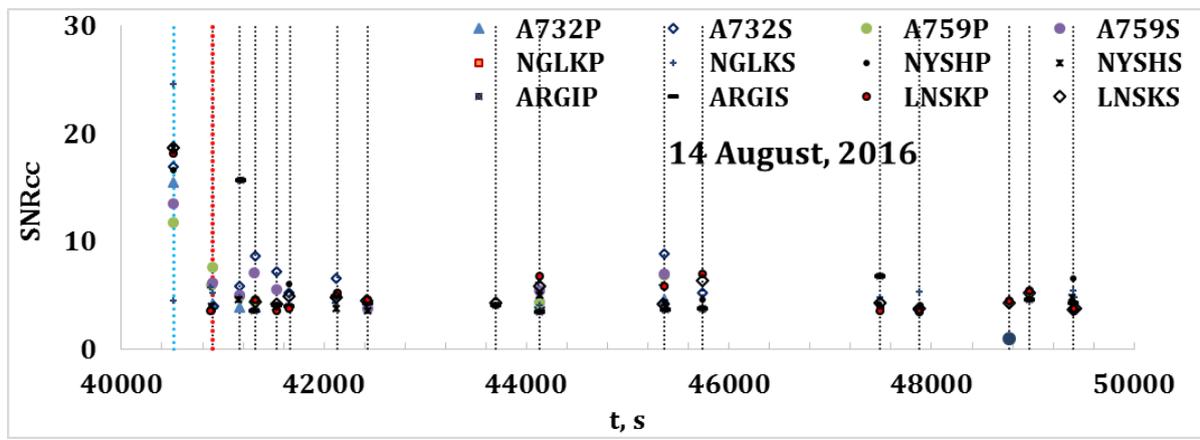

Figure 9. Comparison of the origin times for events from the interactive EQAlert.ru (vertical dotted line) and the XSEL (stations) catalogues. The time segment between 40,000 s and 50,000 s of August 14, 2016 is shown.

Depending on the set of MEs as well as the detection and local association parameters, including those defining the resolution of conflicts between master events, one can obtain aftershock bulletins of varying detail and reliability. Figure 10 compares the magnitude distributions from three catalogues. The interactive catalogue EQAlert.ru is the reference щту. Two cross correlation catalogues were created with the main shock as ME and with 7 MEs. For both XSEL catalogues, the minimum number of associated stations (phases) with one event is 4. As in our previous studies [Bobrov *et al*., 2014; Bobrov, *et al*., 2015ab], the total number of events in XSEL is by 50% to 70% larger than that in the catalogue obtained by standard methods. The gain obtained by 6 MEs added to the main shock is expressed in the increase of the number of events from 194 to 225. A larger number of MEs could increase the number of XSEL events due to their proximity to additional MEs in space and waveform similarity. However, the use of MEs with lower quality may affect the quality of the XSEL catalogue by creation of less reliable event hypotheses.

The frequency distribution of the magnitudes in Figure 10 indicates a slight improvement in the catalogue completeness threshold with the use of 7 MEs, but also indicates an increase in the variation of the relative magnitudes estimates when more than one ME is used. For the main earthquake as a ME, the fall of the distribution above the threshold is of a classical linear nature in the semi-logarithmic coordinates. In the other two catalogues, some deviations from the Gutenberg-Richter law are observed, which are apparently related to the increased uncertainty in amplitude measurement of weak signals. It is important to note that the use of MEs with a small magnitude makes it possible to detect extremely small events with ML <2. In any version, the XSEL contains more events with magnitudes greater than 3 than the catalogue EQAlert.ru. The final confirmation of new hypotheses is possible only in interactive analysis.

For seismotectonic studies, the distribution of aftershocks in space is also important. Location with the WCC and LA methods leads to a better clustering of the aftershocks. In Figure 11, the event locations from the EQAlert.ru and XSEL are compared. In order to further improve the XSEL quality, we increased the minimum number of associated virtual stations to 5, which means association of at least 3 different real stations from 6. The XSEL for this case consists of 170 events listed in Appendix 1. The improved aftershock locations tend to cluster in a narrow zone corresponding to the western (suspended) side of the Central Sakhalin fault, which defines the boundary between the Okhotsk and the Eurasian (Amur) slabs.

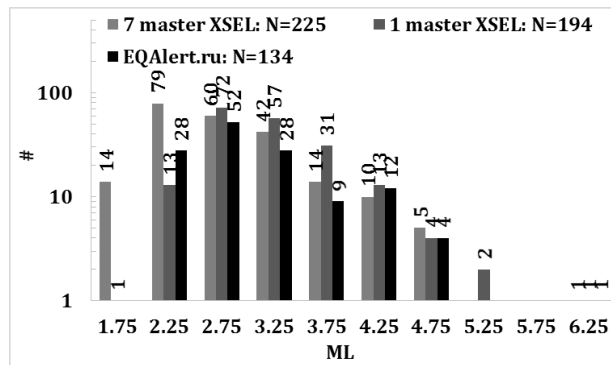

Figure 10. Frequency distribution of magnitudes for three catalogues.

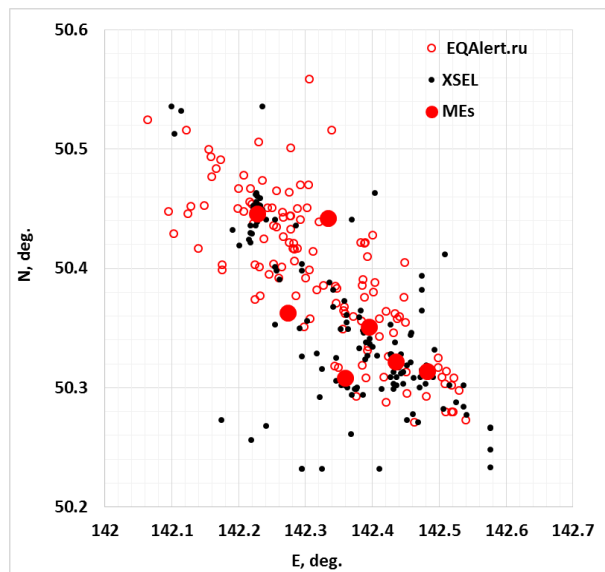

Figure 11. Locations of 133 events from the catalogue EQAlert.ru (red circles) and 170 events from the XSEL (black dots). The latter contains events with at least 3 real stations and 5 virtual stations.

**Discussion**

The WCC method is at the initial stage of its development in general and applied seismology. Despite the profound and comprehensive theoretical justification of the matched filter method in radio-physics as well as in statistics and engineering, there are many aspects of its applications to seismic research related to generation and propagation of signals and to the properties of microseismic noise. Here, we also leave aside the issues of optimal network location and the choice of characteristics of seismometers, such as the comparison of a single three-component station and a small-aperture seismic array consisting of three-component stations.

The results in such previously unexplored with the WCC areas of seismicity, such as Sakhalin Island and its shelf, reveal new empirical relationships and help to clarify the spatial-temporal and magnitude characteristics of seismicity. The cross correlation catalogue contains more events with more accurate estimates of magnitude and location relative to the master events, which are usually better located in absolute terms. New catalogues may serve as a basis for a more accurate assessment of seismicity and the probability of a large earthquake.

Quantitative evaluation of the effectiveness of the WCC method in comparison with existing methods of signal detection and creation of seismic events is based on the comparison of the interactive EQAlert.ru catalogue and the XSEL. The total number of aftershocks found as well as the completeness threshold vary depending on the minimum number of associated stations/phases. In any case, the increase in the number of events found is more than 50%, accompanied by increasing accuracy of location and magnitude estimates. Currently, application of the WCC method is not standard and is limited to several seismic zones, although the interest in the method and the results of its application has been gradually growing.

In general, the seismic network on Sakhalin Island fits the tasks of seismicity monitoring at the threshold of the catalogue completeness ML = 2.5-3. This allows us to solve the main tasks of monitoring regional seismicity and dividing it into natural and man-made. However, with the growth in industrial activity, new processes arise representing a challenge to standard methods of signal detection and event creation.

**Attachment 1.**
Cross correlation catalogue obtained for the minimum number of associated phases 4 (NASS = 4). The time is counted in seconds from the beginning of the day on August 14, 2016. $OT_{RMS}$ is the RMS origin time residual. $OT_{CC}$-$OT_{IMGG}$ is the origin time difference the EQAlert.ru and the XSEL, if they are close. $ML_{RM}$ is the relative magnitude, except for the main event, for which the magnitude is taken from the external catalogue and used as a reference. NASS is the number of associated phases. $CC_{AV}$ is the average value of CC for associated stations. $CC_{CUM}$ is the cumulative value of CC. Master - ME, which found this event. If the XSEL origin time coincides with that in the EQAlert.ru, the last column contains the aftershock serial number.

| Time | E., deg | N., deg | OT$_{RMS}$ | OT$_{CC}$-OT$_{IMGG}$ | ML$_{RM}$ | NASS | CC$_{AV}$ | CC$_{CUM}$ | Master | EQA |
|---|---|---|---|---|---|---|---|---|---|---|
| 40513.0 | 50.351 | 142.395 | 0.003 | -0.09 | **6.100** | 12 | 1 | 12 | 1 | 1 |
| 40771.7 | 50.513 | 142.104 | 0.680 | 0 | 3.022 | 7 | 0.225 | 1.572 | 128 | |
| 40888.4 | 50.365 | 142.382 | 0.597 | -2.63 | 4.388 | 6 | 0.377 | 2.263 | 1 | 3 |
| 41166.7 | 50.308 | 142.359 | 0.003 | -0.09 | 3.800 | 9 | 1 | 9 | 4 | 4 |
| 41289.0 | 50.306 | 142.346 | 0.038 | 0 | 3.202 | 6 | 0.553 | 3.317 | 4 | |
| 41317.1 | 50.312 | 142.482 | 0.015 | 1.02 | 4.248 | 7 | 0.669 | 4.682 | 12 | 5 |
| 41533.6 | 50.319 | 142.481 | 0.024 | -1.5 | 3.035 | 10 | 0.509 | 5.094 | 12 | 6 |
| 41656.4 | 50.284 | 142.536 | 0.400 | -4.43 | 3.070 | 7 | 0.43 | 3.007 | 1 | 7 |
| 41830.5 | 50.309 | 142.359 | 0.006 | 0 | 2.436 | 10 | 0.644 | 6.44 | 4 | |
| 41893.2 | 50.300 | 142.362 | 0.016 | 0 | 2.064 | 5 | 0.476 | 2.378 | 4 | |
| 42122.7 | 50.341 | 142.396 | 0.174 | -4.83 | 2.751 | 6 | 0.401 | 2.406 | 1 | 8 |
| 42275.2 | 50.306 | 142.356 | 0.008 | 0 | 2.193 | 6 | 0.593 | 3.56 | 4 | |
| 42424.8 | 50.337 | 142.396 | 0.808 | -4.97 | 2.905 | 8 | 0.349 | 2.794 | 1 | 9 |
| 43922.6 | 50.303 | 142.479 | 0.259 | 0 | 2.804 | 8 | 0.475 | 3.798 | 12 | |
| 44129.4 | 50.332 | 142.493 | 0.828 | -2.89 | 3.640 | 9 | 0.402 | 3.614 | 82 | 11 |
| 44675.3 | 50.256 | 142.219 | 0.402 | 0 | 2.010 | 6 | 0.396 | 2.379 | 4 | |
| 45359.0 | 50.314 | 142.482 | 0.003 | -0.09 | 3.900 | 9 | 1 | 9 | 12 | 12 |
| 45733.9 | 50.436 | 142.225 | 0.094 | -0.84 | 2.726 | 8 | 0.615 | 4.921 | 128 | 13 |
| 46109.3 | 50.302 | 142.359 | 0.009 | 0 | 1.987 | 5 | 0.547 | 2.735 | 4 | |
| 47486.1 | 50.305 | 142.358 | 0.013 | -3.03 | 2.633 | 11 | 0.621 | 6.83 | 4 | 14 |
| 47873.1 | 50.309 | 142.360 | 0.007 | -3.19 | 2.606 | 11 | 0.804 | 8.839 | 4 | 15 |
| 48159.4 | 50.327 | 142.393 | 0.127 | 0 | 2.085 | 6 | 0.424 | 2.541 | 4 | |
| 48967.2 | 50.424 | 142.215 | 0.016 | -1.59 | 2.395 | 5 | 0.533 | 2.666 | 128 | 17 |
| 49230.8 | 50.299 | 142.374 | 0.178 | 0 | 2.247 | 8 | 0.432 | 3.455 | 4 | |
| 49397.6 | 50.300 | 142.377 | 0.066 | -2.58 | 2.921 | 9 | 0.513 | 4.617 | 4 | 18 |
| 50153.8 | 50.302 | 142.353 | 0.008 | -2.62 | 2.515 | 5 | 0.5 | 2.5 | 4 | 19 |
| 50280.2 | 50.306 | 142.358 | 0.017 | 0 | 2.144 | 7 | 0.521 | 3.645 | 4 | |
| 50839.4 | 50.299 | 142.432 | 0.371 | -1.58 | 2.482 | 8 | 0.489 | 3.911 | 82 | 20 |
| 52559.1 | 50.429 | 142.220 | 0.020 | 0 | 2.228 | 6 | 0.428 | 2.565 | 128 | |
| 52622.7 | 50.430 | 142.218 | 0.023 | 0 | 2.154 | 6 | 0.473 | 2.836 | 128 | |
| 53183.4 | 50.261 | 142.368 | 0.486 | -1.49 | 3.240 | 9 | 0.341 | 3.072 | 1 | 21 |
| 55322.7 | 50.350 | 142.291 | 0.409 | -2.14 | 3.379 | 6 | 0.441 | 2.643 | 4 | 22 |
| 56309.0 | 50.311 | 142.476 | 0.284 | -3.33 | 2.215 | 5 | 0.372 | 1.858 | 4 | 23 |
| 57843.3 | 50.361 | 142.269 | 0.014 | -2.11 | 2.324 | 7 | 0.704 | 4.928 | 129 | 24 |
| 58800.8 | 50.316 | 142.325 | 0.267 | -1.54 | 3.159 | 9 | 0.453 | 4.078 | 4 | 25 |
| 59851.0 | 50.315 | 142.474 | 0.484 | -1.47 | 2.965 | 10 | 0.484 | 4.844 | 12 | 26 |
| 65052.6 | 50.439 | 142.228 | 0.049 | 0 | 1.851 | 6 | 0.394 | 2.366 | 128 | |
| 71503.2 | 50.436 | 142.217 | 0.086 | -1.99 | 2.423 | 7 | 0.567 | 3.966 | 128 | 29 |
| 80485.8 | 50.317 | 142.439 | 0.694 | -3.78 | 3.361 | 8 | 0.397 | 3.177 | 82 | 31 |
| 82159.8 | 50.321 | 142.457 | 0.327 | -4.62 | 2.784 | 7 | 0.405 | 2.836 | 1 | 32 |
| 96812.4 | 50.373 | 142.358 | 1.214 | -3.9 | 2.850 | 6 | 0.363 | 2.176 | 1 | 33 |
| 97345.8 | 50.388 | 142.336 | 0.665 | -3.35 | 3.138 | 7 | 0.373 | 2.61 | 82 | 34 |
| 110382.5 | 50.362 | 142.277 | 0.036 | 0 | 2.592 | 7 | 0.46 | 3.223 | 129 | |
| 116587.6 | 50.344 | 142.457 | 1.026 | 0.19 | 5.264 | 9 | 0.444 | 3.995 | 1 | 36 |
| 118192.7 | 50.232 | 142.294 | 0.366 | 0 | 2.154 | 5 | 0.356 | 1.78 | 82 | |
| 119040.8 | 50.278 | 142.461 | 0.956 | -2.4 | 3.082 | 9 | 0.342 | 3.078 | 1 | 38 |
| 123037.1 | 50.422 | 142.218 | 0.394 | -1.75 | 2.504 | 6 | 0.43 | 2.578 | 128 | 39 |
| 124144.6 | 50.327 | 142.407 | 0.118 | 0 | 2.492 | 5 | 0.397 | 1.987 | 82 | |
| 127665.3 | 50.333 | 142.380 | 0.350 | -2.79 | 3.285 | 7 | 0.442 | 3.093 | 82 | 41 |
| 135307.1 | 50.273 | 142.451 | 0.524 | -0.71 | 2.990 | 8 | 0.387 | 3.098 | 82 | 44 |
| 136902.9 | 50.302 | 142.515 | 1.013 | -0.7 | 2.910 | 6 | 0.326 | 1.954 | 1 | 45 |
| 137711.2 | 50.398 | 142.256 | 1.111 | 0 | 2.362 | 5 | 0.352 | 1.76 | 4 | |
| 140366.8 | 50.441 | 142.369 | 0.755 | 3.12 | 2.564 | 5 | 0.305 | 1.523 | 128 | 46 |
| 153511.8 | 50.232 | 142.294 | 0.927 | 0 | 2.473 | 5 | 0.282 | 1.41 | 82 | |
| 154495.7 | 50.299 | 142.376 | 0.079 | -1.69 | 2.611 | 6 | 0.455 | 2.728 | 4 | 47 |
| 155915.6 | 50.394 | 142.474 | 0.656 | 2.06 | 3.721 | 6 | 0.355 | 2.129 | 133 | 48 |
| 157232.0 | 50.313 | 142.487 | 1.268 | -0.22 | 3.536 | 8 | 0.374 | 2.994 | 82 | 49 |
| 178589.6 | 50.355 | 142.361 | 0.910 | -2.72 | 3.158 | 6 | 0.329 | 1.974 | 1 | 51 |
| 180319.7 | 50.401 | 142.254 | 0.847 | -4.92 | 2.968 | 5 | 0.38 | 1.901 | 1 | 52 |
| 183612.7 | 50.356 | 142.302 | 1.341 | -1.07 | 3.919 | 9 | 0.37 | 3.334 | 1 | 53 |
| 184578.3 | 50.324 | 142.389 | 0.639 | -1.34 | 4.258 | 9 | 0.454 | 4.089 | 1 | 55 |
| 188706.2 | 50.368 | 142.341 | 1.049 | -0.42 | 4.065 | 7 | 0.372 | 2.602 | 12 | 56 |
| 190627.4 | 50.463 | 142.404 | 0.531 | -0.17 | 4.417 | 5 | 0.436 | 2.181 | 133 | 59 |
| 196338.6 | 50.453 | 142.228 | 0.036 | 0 | 2.612 | 6 | 0.567 | 3.404 | 128 | |
| 203460.3 | 50.382 | 142.341 | 0.375 | 0 | 2.565 | 5 | 0.369 | 1.843 | 82 | |
| 203910.8 | 50.449 | 142.235 | 0.070 | 0.04 | 3.186 | 7 | 0.445 | 3.116 | 128 | 60 |
| 204609.2 | 50.451 | 142.232 | 0.092 | 0 | 2.731 | 7 | 0.402 | 2.816 | 128 | |
| 207439.3 | 50.312 | 142.443 | 0.452 | 0.03 | 3.198 | 8 | 0.43 | 3.442 | 12 | 61 |
| 222835.3 | 50.294 | 142.386 | 1.126 | -1.44 | 3.966 | 9 | 0.473 | 4.253 | 82 | 63 |
| 229143.9 | 50.432 | 142.191 | 0.596 | 0 | 2.190 | 6 | 0.353 | 2.119 | 128 | |
| 238014.6 | 50.354 | 142.399 | 0.721 | -2.93 | 2.897 | 8 | 0.35 | 2.796 | 1 | 64 |
| 238114.5 | 50.436 | 142.285 | 0.223 | 0 | 2.207 | 5 | 0.373 | 1.866 | 128 | |
| 242498.1 | 50.391 | 142.261 | 0.798 | -2.76 | 3.261 | 7 | 0.332 | 2.325 | 1 | 65 |
| 249343.2 | 50.361 | 142.361 | 0.983 | -1.36 | 4.826 | 10 | 0.434 | 4.34 | 1 | 66 |
| 249495.2 | 50.446 | 142.228 | 0.011 | 0 | 4.060 | 10 | 0.867 | 8.669 | 128 | |
| 249608.3 | 50.453 | 142.222 | 0.007 | 0 | 3.138 | 8 | 0.561 | 4.485 | 128 | |
| 250199.7 | 50.442 | 142.228 | 0.016 | 0 | 3.113 | 9 | 0.809 | 7.277 | 128 | 67 |
| 252133.6 | 50.446 | 142.228 | 0.000 | 0 | 2.213 | 5 | 0.596 | 2.978 | 128 | |
| 255216.1 | 50.267 | 142.576 | 0.905 | 0.32 | 3.422 | 7 | 0.417 | 2.917 | 82 | 69 |
| 255381.0 | 50.452 | 142.229 | 0.033 | 0 | 2.755 | 7 | 0.577 | 4.036 | 128 | |
| 255917.2 | 50.449 | 142.232 | 0.019 | 0 | 2.587 | 6 | 0.51 | 3.061 | 128 | |
| 257883.4 | 50.233 | 142.576 | 1.407 | -1.09 | 4.090 | 7 | 0.451 | 3.158 | 82 | 70 |
| 258394.4 | 50.359 | 142.380 | 0.219 | -3.49 | 2.979 | 5 | 0.374 | 1.869 | 82 | 71 |
| 259958.2 | 50.450 | 142.229 | 0.008 | -1.9 | 2.843 | 7 | 0.652 | 4.562 | 128 | 72 |
| 260351.5 | 50.447 | 142.228 | 0.008 | 0 | 2.471 | 7 | 0.575 | 4.026 | 128 | |
| 261443.3 | 50.446 | 142.227 | 0.005 | 0 | 2.451 | 6 | 0.595 | 3.571 | 128 | |
| 263566.2 | 50.349 | 142.363 | 0.394 | -3.21 | 2.969 | 5 | 0.335 | 1.674 | 82 | 73 |
| 263855.3 | 50.382 | 142.474 | 0.784 | -3.33 | 3.273 | 6 | 0.37 | 2.22 | 133 | 74 |
| 269207.1 | 50.302 | 142.536 | 1.220 | 0 | 2.634 | 5 | 0.339 | 1.695 | 1 | |
| 269402.1 | 50.300 | 142.470 | 0.260 | -2.5 | 3.442 | 10 | 0.603 | 6.031 | 82 | 76 |
| 274670.7 | 50.319 | 142.451 | 0.218 | -4.08 | 2.831 | 7 | 0.483 | 3.379 | 82 | 77 |
| **279044.2** | 50.453 | 142.225 | 0.025 | 0.38 | 2.548 | 7 | 0.562 | 3.932 | 128 | 78 |
| **293672.8** | 50.303 | 142.446 | 0.163 | -2.47 | 3.174 | 9 | 0.558 | 5.019 | 82 | 79 |
| **305401.5** | 50.441 | 142.254 | 1.324 | -4.23 | 3.461 | 8 | 0.382 | 3.055 | 1 | 80 |
| **305860.9** | 50.353 | 142.427 | 0.238 | 0 | 2.277 | 6 | 0.372 | 2.233 | 82 | |
| **308378.5** | 50.307 | 142.352 | 0.024 | -1.42 | 2.892 | 8 | 0.665 | 5.32 | 4 | 81 |
| **309797.5** | 50.322 | 142.435 | 0.000 | -0.1 | 4.900 | 11 | 1 | 11 | 82 | 82 |
| **309877.0** | 50.448 | 142.227 | 0.007 | 0 | 3.119 | 6 | 0.558 | 3.347 | 128 | |
| **310111.3** | 50.338 | 142.390 | 0.203 | 0 | 2.937 | 7 | 0.536 | 3.754 | 82 | |
| **310425.6** | 50.325 | 142.439 | 0.334 | -3.41 | 3.107 | 8 | 0.57 | 4.558 | 82 | 83 |
| **311160.5** | 50.451 | 142.222 | 0.012 | 0.46 | 2.202 | 8 | 0.544 | 4.348 | 128 | 84 |
| **311276.2** | 50.277 | 142.541 | 0.414 | -0.71 | 3.808 | 8 | 0.583 | 4.663 | 82 | 85 |
| **311541.9** | 50.232 | 142.324 | 0.731 | 0 | 2.324 | 6 | 0.477 | 2.861 | 82 | |
| **311601.1** | 50.232 | 142.410 | 0.176 | -0.27 | 2.472 | 5 | 0.479 | 2.397 | 82 | 86 |
| **312132.2** | 50.266 | 142.576 | 0.596 | -2.98 | 2.429 | 9 | 0.472 | 4.251 | 82 | 87 |
| **312240.5** | 50.327 | 142.428 | 0.300 | -3.06 | 2.618 | 9 | 0.464 | 4.179 | 82 | 88 |
| **313384.5** | 50.313 | 142.446 | 0.250 | -3.44 | 2.488 | 8 | 0.437 | 3.496 | 82 | 89 |
| **314292.0** | 50.309 | 142.436 | 0.245 | 0 | 2.134 | 6 | 0.461 | 2.764 | 82 | |
| **316232.2** | 50.271 | 142.468 | 0.341 | 1.15 | 4.661 | 7 | 0.443 | 3.101 | 12 | 90 |
| **317235.3** | 50.302 | 142.436 | 0.338 | -2.22 | 3.032 | 9 | 0.512 | 4.608 | 82 | 91 |
| **317342.4** | 50.456 | 142.227 | 0.039 | 0 | 2.175 | 7 | 0.493 | 3.452 | 128 | |
| **317544.2** | 50.308 | 142.462 | 0.139 | -1.98 | 3.243 | 9 | 0.599 | 5.387 | 82 | 92 |
| **319408.2** | 50.321 | 142.443 | 0.023 | 0 | 2.324 | 7 | 0.517 | 3.616 | 82 | |
| **322609.3** | 50.328 | 142.443 | 0.061 | 0 | 1.857 | 5 | 0.382 | 1.909 | 82 | |
| **322951.6** | 50.412 | 142.508 | 0.287 | 0 | 1.683 | 5 | 0.351 | 1.757 | 82 | |
| **328199.4** | 50.461 | 142.228 | 0.011 | -5.64 | 2.287 | 5 | 0.409 | 2.047 | 128 | 93 |
| **331214.7** | 50.346 | 142.387 | 0.209 | 0 | 2.339 | 6 | 0.389 | 2.336 | 82 | |
| **333204.9** | 50.365 | 142.474 | 0.675 | 0 | 3.216 | 6 | 0.324 | 1.945 | 1 | |
| **341952.6** | 50.347 | 142.387 | 0.220 | 0 | 2.770 | 6 | 0.475 | 2.849 | 82 | |
| **342499.5** | 50.309 | 142.427 | 0.072 | 0 | 2.469 | 7 | 0.496 | 3.475 | 82 | |
| **343309.1** | 50.328 | 142.428 | 0.193 | 0 | 2.359 | 5 | 0.364 | 1.822 | 82 | |
| **348078.7** | 50.309 | 142.490 | 0.495 | -0.22 | 3.619 | 11 | 0.46 | 5.061 | 12 | 94 |
| **348636.9** | 50.448 | 142.228 | 0.015 | -1.86 | 2.771 | 9 | 0.715 | 6.439 | 128 | 95 |
| **350410.1** | 50.398 | 142.294 | 1.234 | -2.89 | 3.172 | 8 | 0.331 | 2.649 | 82 | 96 |
| **355699.7** | 50.364 | 142.269 | 0.013 | 0 | 2.778 | 8 | 0.708 | 5.663 | 129 | |
| **368922.6** | 50.282 | 142.506 | 0.929 | -0.32 | 3.042 | 6 | 0.412 | 2.475 | 12 | 98 |
| **370760.9** | 50.448 | 142.228 | 0.020 | -2.53 | 2.604 | 7 | 0.682 | 4.774 | 128 | 99 |
| **378433.8** | 50.462 | 142.225 | 0.091 | -1.49 | 2.861 | 8 | 0.526 | 4.207 | 128 | 100 |
| **383156.8** | 50.453 | 142.232 | 0.004 | 0 | 2.147 | 5 | 0.436 | 2.182 | 128 | |
| **384797.1** | 50.445 | 142.228 | 0.007 | 0 | 2.506 | 8 | 0.668 | 5.345 | 128 | |
| **387283.0** | 50.309 | 142.472 | 0.149 | 0 | 2.433 | 5 | 0.359 | 1.795 | 82 | |
| **393253.2** | 50.448 | 142.221 | 0.009 | 0 | 2.231 | 6 | 0.489 | 2.934 | 128 | |
| **394855.4** | 50.446 | 142.227 | 0.012 | 0 | 2.222 | 7 | 0.575 | 4.025 | 128 | |
| **395246.3** | 50.326 | 142.294 | 0.809 | -3.41 | 2.437 | 6 | 0.389 | 2.333 | 82 | 101 |
| **396930.2** | 50.441 | 142.241 | 0.007 | 0 | 2.080 | 5 | 0.523 | 2.616 | 128 | |
| **398504.3** | 50.463 | 142.227 | 0.075 | -1.88 | 2.495 | 6 | 0.481 | 2.884 | 128 | 102 |
| **400196.1** | 50.328 | 142.427 | 0.140 | -3.27 | 2.987 | 9 | 0.515 | 4.636 | 82 | 103 |
| **404023.2** | 50.288 | 142.525 | 0.613 | -2.88 | 3.461 | 8 | 0.422 | 3.373 | 1 | 104 |
| **405721.7** | 50.536 | 142.235 | 0.762 | 0 | 2.308 | 5 | 0.318 | 1.592 | 128 | |
| **413418.4** | 50.310 | 142.358 | 0.010 | -3.76 | 2.666 | 10 | 0.697 | 6.972 | 4 | 105 |
| **431972.8** | 50.353 | 142.254 | 1.290 | 0 | 4.137 | 5 | 0.275 | 1.374 | 1 | |
| **436790.7** | 50.338 | 142.434 | 0.860 | -0.91 | 4.557 | 8 | 0.426 | 3.409 | 1 | 106 |
| **448930.7** | 50.248 | 142.576 | 1.182 | -1.28 | 2.982 | 6 | 0.375 | 2.247 | 82 | 109 |
| **490737.0** | 50.536 | 142.099 | 0.945 | -0.34 | 2.219 | 5 | 0.359 | 1.793 | 128 | 112 |
| **492889.3** | 50.303 | 142.359 | 0.015 | -3.63 | 2.099 | 6 | 0.478 | 2.87 | 4 | 113 |
| **495835.3** | 50.532 | 142.114 | 0.428 | 0 | 1.973 | 5 | 0.323 | 1.615 | 128 | |
| **515454.1** | 50.446 | 142.227 | 0.010 | 0 | 2.210 | 5 | 0.651 | 3.254 | 128 | |
| **526033.8** | 50.292 | 142.321 | 0.950 | 0.14 | 4.354 | 10 | 0.441 | 4.413 | 4 | 115 |
| **528479.8** | 50.334 | 142.400 | 0.189 | 0 | 2.346 | 6 | 0.413 | 2.48 | 82 | |
| **545440.3** | 50.329 | 142.317 | 1.193 | -4.36 | 3.589 | 7 | 0.386 | 2.701 | 1 | 117 |
| **549942.3** | 50.313 | 142.431 | 0.195 | -3.25 | 2.685 | 8 | 0.433 | 3.462 | 82 | 118 |
| **553921.5** | 50.299 | 142.414 | 0.379 | -2.28 | 2.743 | 5 | 0.394 | 1.971 | 82 | 119 |
| **556863.8** | 50.445 | 142.227 | 0.007 | 0 | 2.057 | 6 | 0.552 | 3.314 | 128 | |
| **561595.5** | 50.346 | 142.458 | 0.151 | -4.25 | 2.344 | 8 | 0.418 | 3.346 | 82 | 120 |
| **569925.0** | 50.305 | 142.355 | 0.027 | -2.49 | 2.810 | 11 | 0.711 | 7.824 | 4 | 121 |
| **638731.3** | 50.404 | 142.294 | 1.028 | 0 | 2.132 | 5 | 0.377 | 1.885 | 82 | |
| **711443.3** | 50.325 | 142.346 | 0.371 | -3.54 | 3.468 | 6 | 0.406 | 2.435 | 82 | 122 |
| **739626.8** | 50.303 | 142.432 | 0.379 | -2.87 | 2.771 | 9 | 0.456 | 4.1 | 82 | 123 |
| **750172.3** | 50.268 | 142.241 | 0.733 | -1.57 | 2.560 | 5 | 0.326 | 1.628 | 4 | 124 |
| **817739.1** | 50.348 | 142.386 | 0.186 | -1.38 | 2.539 | 8 | 0.451 | 3.608 | 82 | 125 |
| **848486.7** | 50.459 | 142.232 | 0.021 | 0 | 2.172 | 5 | 0.473 | 2.364 | 128 | |
| **1086780.6** | 50.349 | 142.352 | 0.751 | -2.6 | 2.523 | 6 | 0.401 | 2.408 | 82 | 126 |
| **1125693.1** | 50.419 | 142.201 | 0.214 | -0.94 | 2.811 | 5 | 0.407 | 2.034 | 128 | 127 |
| **1149831.1** | 50.446 | 142.228 | 0.000 | 0.03 | 3.300 | 10 | 1 | 10 | 128 | 128 |
| **1194263.0** | 50.363 | 142.273 | 0.000 | 0 | 4.700 | 9 | 1 | 9 | 129 | 129 |
| **1195112.3** | 50.361 | 142.270 | 0.014 | -1.69 | 3.051 | 9 | 0.747 | 6.725 | 129 | 130 |
| **1235545.4** | 50.362 | 142.279 | 0.052 | 0.42 | 3.758 | 9 | 0.674 | 6.063 | 129 | 131 |
| **1252030.2** | 50.273 | 142.174 | 0.207 | 0 | 2.061 | 5 | 0.343 | 1.716 | 129 | |
| **1281743.8** | 50.294 | 142.369 | 0.041 | -3.01 | 2.537 | 6 | 0.426 | 2.556 | 4 | 132 |
| **1367513.3** | 50.442 | 142.333 | 0.000 | 0.05 | 4.100 | 5 | 1 | 5 | 133 | 133 |